\documentclass[a4paper,11pt]{article}
\pdfoutput=1 

\usepackage{jcappub} 
                     
\usepackage[T1]{fontenc} 

\usepackage{bm}
\usepackage{gensymb}

\usepackage{graphicx}
\usepackage{subcaption}

\usepackage{color}

\usepackage{epstopdf}

\graphicspath{{Images/}}

\def \ru{\hat{\bm{r}}}
\def \phiu{\hat{\bm{\phi}}}

\title{\boldmath Deflections of ultra-high energy cosmic rays by the Milky Way magnetic field: how well can they be corrected?}

 \author[a]{G. Magkos}
 \author[a,b,c]{and V. Pavlidou}
 \affiliation[a]{Department of Physics, University of Crete,\\71003, Heraklion, Greece}
 \affiliation[b]{Foundation for Research and Technology - Hellas, IESL and Institute of Astrophysics, \\71110, Heraklion, Greece}
 \affiliation[c]{Institute for Theoretical and Computational Physics, University of Crete,\\71003, Heraklion, Greece}

\emailAdd{gmagkos@physics.uoc.gr}
\emailAdd{pavlidou@physics.uoc.gr}

\abstract{
Locating the sources of ultra-high energy cosmic rays (UHECRs) still remains a difficult puzzle for modern astrophysics. A major hurdle in the search for the sources is the fact that UHECRs are deflected by the Galactic magnetic field (GMF). Current knowledge of the GMF is limited, as most experimental measurements track line-of-sight--integrated quantities that are used to obtain best-fit parameters for global models including a large random component. The advent, however, of the Gaia era, with measurements of $\sim 10^9$ stellar parallaxes, in combination with upcoming large polarimetric surveys, make, for the first time, a 3D measurement of the GMF possible in principle. Such measurements can then be used to attempt a reconstruction of the trajectories of individual UHECRs through the Galaxy, in order to correct for their deflection.

Motivated by these developments, in the present work, we study the limits of such a correction, by examining how its effectiveness depends on the uncertainty of any such future magnetic field measurements. To that end, we simulate attempts to reconstruct the trajectory of the cosmic ray by using hypothetical measurements of the GMF, based on values received from two recently updated GMF models. To simulate the uncertainty of a 3D measurement, random errors to these values are introduced separately for the plane-of-the-sky (POS) magnitude, the line-of-sight (LOS) magnitude and the POS direction.

Our results highlight the conditions under which an effective correction is achievable. We find that the effectiveness is dependent on the particle rigidity and arrival direction, and can vary significantly depending on the GMF model used. 	
}

\begin{document}

\maketitle
\flushbottom

\section{Introduction}
\label{sec:intro}

Ultra high energy cosmic rays (UHECRs) are the most energetic particles ever detected, with energies in some cases exceeding $10^{20} \  {\rm eV}$. Their sources, mechanism of acceleration, and composition are currently unknown and debated \cite{Aloisio}. One reason for the difficulty to reach safe conclusions in any of the above areas is the extremely small flux of the incoming UHECRs, which leads to poor statistics. For example, for energies of about $10^{20} \ {\rm eV}$, the flux is of the order of one particle per ${\rm km}^2$ per century \cite{TA}. Another reason is the fact that UHECRs, consisting of protons or heavier nuclei, therefore charged particles, are deflected by the Galactic magnetic field (GMF) as well as the intergalactic magnetic field. In the present work, we focus on the role of the GMF.

The main limitation in studies of the GMF is that currently available observables that probe the GMF are integrated along the line of sight (LOS). Such observables include Faraday rotation measures, synchrotron intensity and polarization of dust emission (see, e.g., \cite{Han} for a recent review on magnetic field observations). In the absence of 3D tomographic information, the approach that has been used for the mapping of the GMF is parameter fitting of different GMF components, including a large random component (e.g., \cite{JF12_ord,JF12_rand,Sun8,Sun10,Planck}). 

However, the advent of the Gaia mission \cite{Gaia}, which is expected to provide parallaxes for a, for the first time, significant fraction of the Galactic stellar population, is redefining our ability to conduct tomographic studies of the Galaxy including its magnetic field. Gaia stellar distances, combined with optopolarimetric measurements of starlight, will provide a   unique and previously unavailable handle on the GMF structure. As dust grains in interstellar clouds tend to align themselves perpendicular to the direction of the magnetic field, optical light passing through a cloud will be absorbed preferentially in the direction of the grain long axis and will be polarized in the plane-of-the-sky (POS) direction of the magnetic field there \cite{Dust}. Upcoming optopolarimetric surveys such as PASIPHAE \cite{PASIPHAE, PASIPHAE_paper} are expected to provide a large number ($>10^6$) of high-quality stellar polarization measurements that can yield 3D information for the POS direction (measured directly) as well as the POS magnitude of the GMF (inferred using e.g. the Davis-Chandrasekhar-Fermi method \cite{Davis,ChF,Hildebrand,Panopoulou,Soam_Pattle,Liu,Beuther,Kwon,Mao,Soam_Lee,Clemens}). 

Any such 3D measurements of the GMF could be used to reconstruct the trajectories of individual cosmic rays through the Galaxy, in order to obtain the original arrival direction of a cosmic ray before it was deflected by the GMF. The effectiveness of such a correction would depend on the accuracy of any such future measurement. If an effective correction were possible in this way, it would constitute an important development in cosmic-ray astronomy, as it could result in increased clustering of cosmic-ray arrival directions, and improved estimates for the location of sources.

In the present work, we study the limits of this method of trajectory reconstruction for individual UHECRs, independently of any specific dataset, experiment, or technique for the determination of the magnetic field direction and strength. We simulate this process by using hypothetical 3D measurements of the GMF. We examine how the precision of the measurements can affect the quality of the reconstruction and how this quality is affected by the rigidity (i.e. the energy of the particle divided by its charge) and arrival direction of the cosmic ray at the Earth. This is accomplished as follows. First, given the rigidity and arrival direction, we find the ``true'' path of the cosmic ray by using a specific GMF model to provide the ``true''  values of the GMF. We then obtain the original arrival direction of the cosmic ray before it entered the Galaxy. Next, we introduce random errors to the ``true'' GMF values along the trajectory in order to simulate the uncertainty of some hypothetical GMF measurements. Again, we find the estimated cosmic ray trajectory using the GMF values with errors and obtain the estimated original arrival direction. Finally, we quantify the effectiveness of our correction by calculating how close the estimated original arrival direction is to the ``true'' original arrival direction.

\section{Methods}
\label{sec:Methods}

\subsection{Propagation of an UHECR in a constant and uniform magnetic field}

An UHECR, thought to be a proton or a nucleus, has charge $q=Z e$, where $Z$ is the number of protons in the nucleus and $e$ the charge of a proton. As it is an ultra-relativistic particle of $\gamma \sim 10^{9}$--$10^{10}$ for $E \sim 10^{19} \  {\rm eV}$, we can safely neglect its rest mass and consider its speed to be equal to the speed of light $c$. 

The velocity $\bm{v}$ of a relativistic charge $q$ with energy $E$ in a constant and uniform magnetic field $\bm{B}$ follows:

\begin{equation}
\dfrac{d\bm{v}}{dt}=\dfrac{c^2 q}{E} (\bm{v} \times \bm{B} ).
\label{eq:velocity}
\end{equation}

The quantity that defines how much the cosmic ray is deflected in a given magnetic field is the rigidity $R=E/q=E/Z e$. Using eq. \ref{eq:velocity} for small time intervals $\delta t$ we can numerically calculate the evolution of the velocity and position of the particle. The time interval that we use is such that $c \, \delta t = 1 \, {\rm pc}$, which is much smaller than the typically assumed coherence length $\sim 100 \, {\rm pc}$ of the GMF  (e.g. \cite{Golup}).

\subsection{Reconstruction of the ``true'' cosmic ray trajectory}

To reconstruct the cosmic ray trajectory, we construct, within the Galaxy, a cubic grid with side $L=100 \  {\rm pc}$. We assume that the magnetic field is uniform inside each cube. The length $L$ has been chosen to be similar to the coherence length of the GMF. 

Before the trajectory reconstruction begins, the magnetic field for each grid cube is calculated using a GMF model and our result is assumed to be the ``true'' value within that cube. We use two different GMF models, so as to have an indication of the influence that the choice of GMF model has on our results. The models used are the Jansson \& Farrar model (hereafter JF12) \cite{JF12_ord,JF12_rand} and the Sun \& Reich model (hereafter Sun10) \cite{Sun8,Sun10}, as updated recently by the Planck collaboration \cite{Planck}. Details on the models and parameter values are given in appendix \ref{sec:GMF}. 

Both models, besides a regular component, include a random component, which is not explicitly specified. The only information given is its RMS value $B_{RMS} (\bm{r})$ as a function of the position. It is considered isotropic, i.e. it has equal probability to point in any direction. We model the random component using three random gaussian variables with a mean of 0 and a standard deviation of $B_{RMS}/\sqrt{3}$, representing the x, y and z components of the magnetic field. Note that this method does not enforce the requirement $\nabla \cdot \bm{B} = 0$ for the random field. A more detailed way to simulate the random field would be to use a Kolmogorov random field, as  explained in \cite{Keivani_diss} and implemented in an updated version of the CRT numerical propagation code \cite{CRT}. However, this choice in implementation is expected to affect our results negligibly compared to the much more significant uncertainties in the overall GMF geometry itself, as we will see in section \ref{sec:Results}.

After calculating the magnetic field, we let time run backwards in steps of $c \, \delta t$ and, using eq. (\ref{eq:velocity}), we calculate for each step the position and velocity of the cosmic ray, as it propagates in the cubic grid. The only information needed to reconstruct the cosmic ray trajectory, given the magnetic field, is the rigidity and arrival direction of the cosmic ray.

Eventually, as time runs backwards, the cosmic ray reaches a point where the strength of the GMF is negligible, and so there is no more Galaxy-induced deflection. The choice of a specific point to place such a cutoff is somewhat arbitrary. We have chosen to place our cutoff at a height of 10 kpc north or south of the Galactic plane and at a radius (in cylindrical coordinates) of 20 kpc from the Galactic center. 

When the cosmic ray reaches this cutoff, we obtain the direction of its velocity vector, from which we infer the ``true'' original arrival direction of the particle before it entered the Galaxy. If the particle has not suffered significant deflections during its propagation in the intergalactic medium, this direction should be close to the direction of its source.

\subsection{Reconstruction of the cosmic ray trajectory with uncertainties}

\label{subsec:errors}

Having found the ``true'' trajectory and original arrival direction of the cosmic ray, we then simulate attempts to reconstruct the trajectory given hypothetical measurements of the GMF with some uncertainty. We perform this by repeating the above process of reconstruction, but this time introducing random errors to the ``true'' magnetic field values found using a GMF model. When the cosmic ray reaches a new grid cube during its backward propagation, the ``measured'' value of the magnetic field for that cube is calculated. 

Errors are introduced separately for the POS magnitude, the LOS magnitude and the POS direction of the magnetic field. For the POS and LOS magnitudes, the ``measured'' values of the respective magnetic field component are considered to follow a lognormal distribution. The mean of this distribution is placed at the ``true'' value $B$ of the respective component received from the GMF model, and the standard deviation $\sigma_B$ is proportional to $B$ so that the ratio $\sigma_B/B$ stays the same for all grid cubes.
This is implemented by multiplying the ``true'' value $B$ by a random lognormal variable $r$ with a mean of 1 and standard deviation $\sigma_r = \sigma_B/B$. For the POS direction, the ``true'' POS direction is rotated by a random angular variable that follows a von Mises distribution with a mean of zero and different values of the parameter $1/\sqrt{\kappa}$ (see appendix \ref{sec:vonMises}), which, for large $\kappa$, approaches the standard deviation of a normal distribution.

Again, we obtain the direction of the velocity of the cosmic ray when it reaches the cutoff, from which we can infer the estimated original arrival direction. We then repeat the above process for a large number of iterations with different random errors.

\subsection{Quality of the reconstruction}

To quantify how well the cosmic ray trajectory has been reconstructed given the uncertainty of our hypothetical GMF measurements, we calculate the post-correction residual angle $\phi_{res}$ between the ``true'' and the estimated original arrival direction. We also calculate the total deflection angle $\phi_{defl}$ of the cosmic ray, as the angle between the ``true'' original arrival direction and the arrival direction at the Earth. We then compare the two angles, introducing the ``effectiveness coefficient'' 

\begin{equation}
a=\frac{\phi_{res}}{\phi_{defl}} .
\end{equation}
This coefficient measures the effectiveness of our correction. Smaller values of $a$ mean a better correction. A value of $a=1$ means that the residual angle is equal to the deflection, and thus the correction is not helpful. A value of $a>1$ means that the residual angle is even larger than the deflection angle, which means that the correction has actually made matters worse.  

During each iteration with different random errors, we collect the values of $\phi_{res}$ and $a$ to find their distribution. We do this for different choices of uncertainties and different initial conditions (arrival direction and rigidity). 

\subsection{Validation}

We have tested our code for the simple case of a uniform magnetic field. Further, we have successfully reproduced the deflection map for the regular field of the original JF12 model (Figure 11 of \cite{JF12_ord}), using, only in this occasion, the original parameters without the update from the Planck collaboration. 

Additionally, we have performed a test of how the discretization of the magnetic field in cubes of uniform field with side 100 pc can affect the simulated propagation of the cosmic ray. Using the regular field of the (updated) JF12 model, we have simulated the propagation of a cosmic ray using three different discretization scenarios for the magnetic field. The first scenario is the same as before, using cubes of side 100 pc containing a uniform field. For the second scenario, we have doubled the cube side to 200 pc, expecting that this would make possible effects of the discretization more apparent. Finally, for the third scenario we did not use a cubic grid, but used the value for the magnetic field at the exact location of the cosmic ray during each 1-pc step of the numerical propagation, essentially using an almost continuous field. 
	
We performed this test using 60 EeV particles (either protons or iron nuclei) for 10 arrival directions. To compare between the results of the three scenarios, we calculated the difference between the total deflections experienced by the cosmic ray in each scenario. The resulting differences between the first and third scenarios were usually less than 1\% and the largest difference found was $\sim 3\%$. The differences between the second and third scenarios were in general larger, but still smaller than $\sim 6\%$, except for a single value that was at $\sim 21\%$.
\footnote{The 21\% difference was found for an arrival direction with Galactic longitude $l=90\degree$ and Galactic latitude $b=-40\degree$, assuming iron composition, which in general means larger deflections than protons, and thus larger differences between scenarios are to be expected.}
These results suggest that using cubes of side 100 pc containing a uniform magnetic field does not have a significant effect on the propagation, while using larger cubes could have a more pronounced effect.

\section{Results}

\label{sec:Results}

\subsection{Effectiveness Coefficient}

\begin{figure}[t]  
\includegraphics[width=\textwidth]{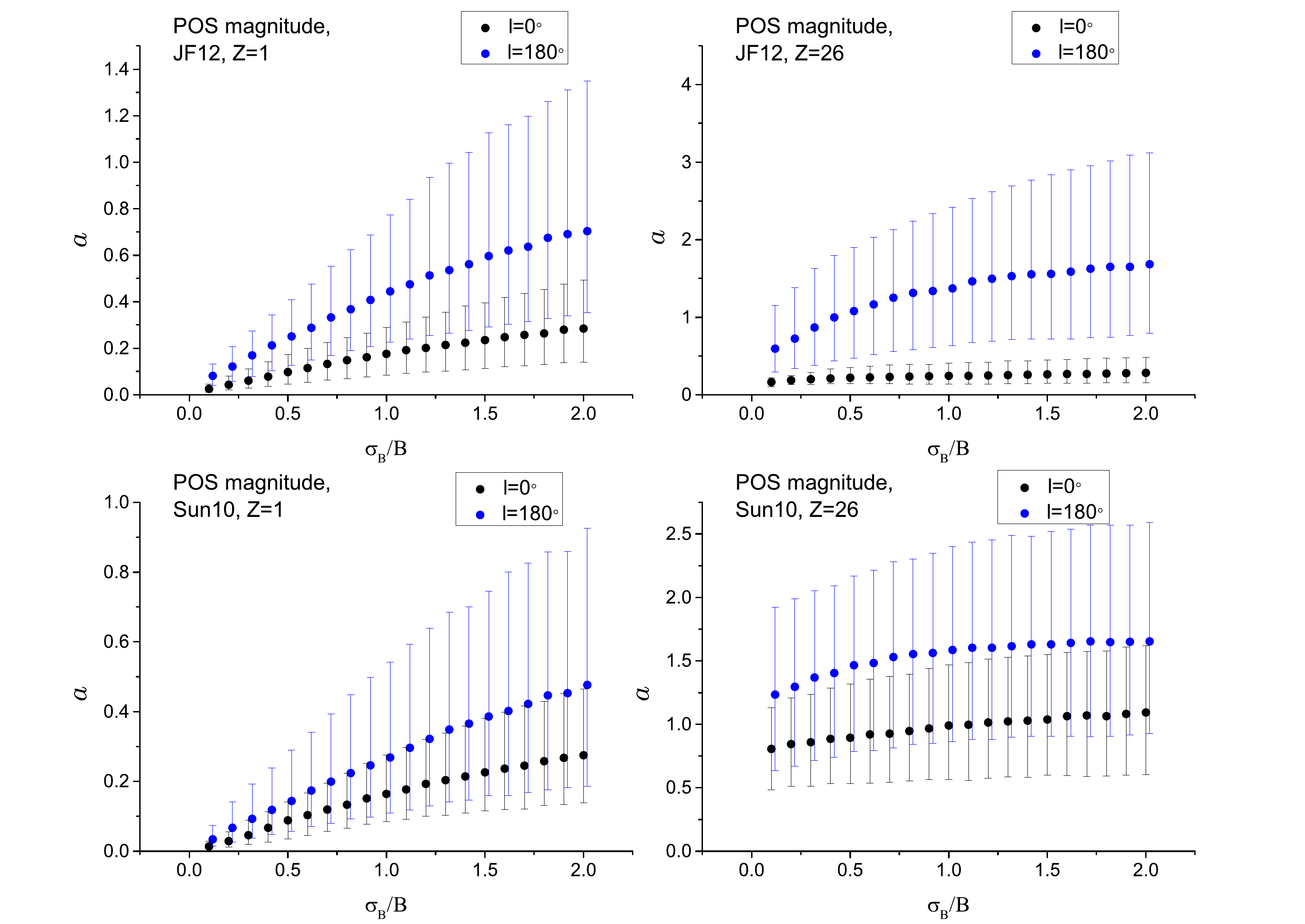}
\caption{Median and 68\% confidence interval of the effectiveness coefficient $a$ as a function of the uncertainty $\sigma_B/B$ of the POS magnitude of the magnetic field for different GMF models, nuclei and Galactic longitudes $l$. Particles of energy 60 EeV and Galactic latitude $b=40\degree$ are assumed. Lower values of $a$ indicate a better correction. }
\label{fig:a_POSmagn}
\end{figure}

\begin{figure}[t]  
\includegraphics[width=\textwidth]{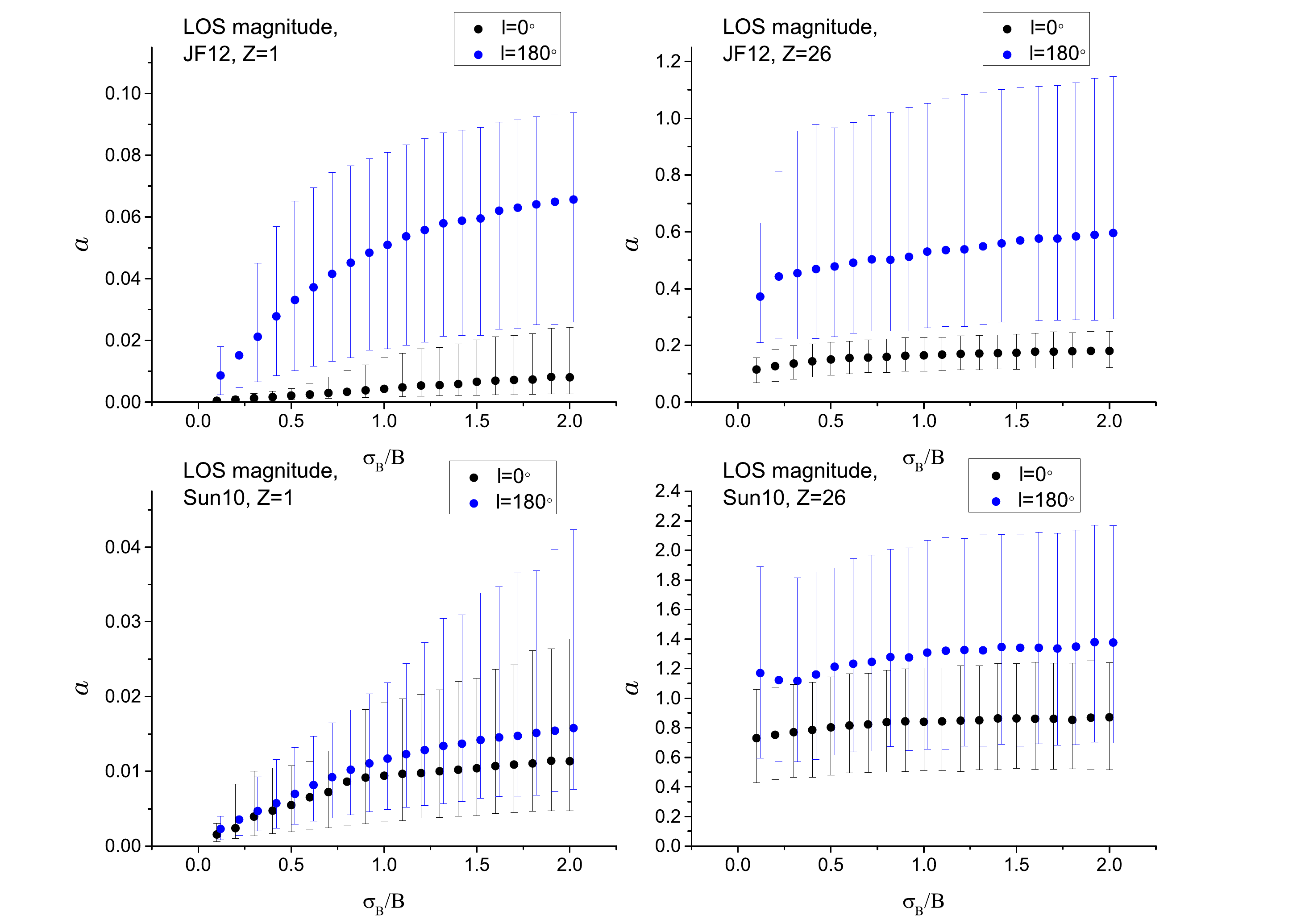}
\caption{As in figure \ref{fig:a_POSmagn}, but for the uncertainty $\sigma_B/B$ of the LOS magnitude.}
\label{fig:a_LOSmagn}
\end{figure}

\begin{figure}[t]  
\includegraphics[width=\textwidth]{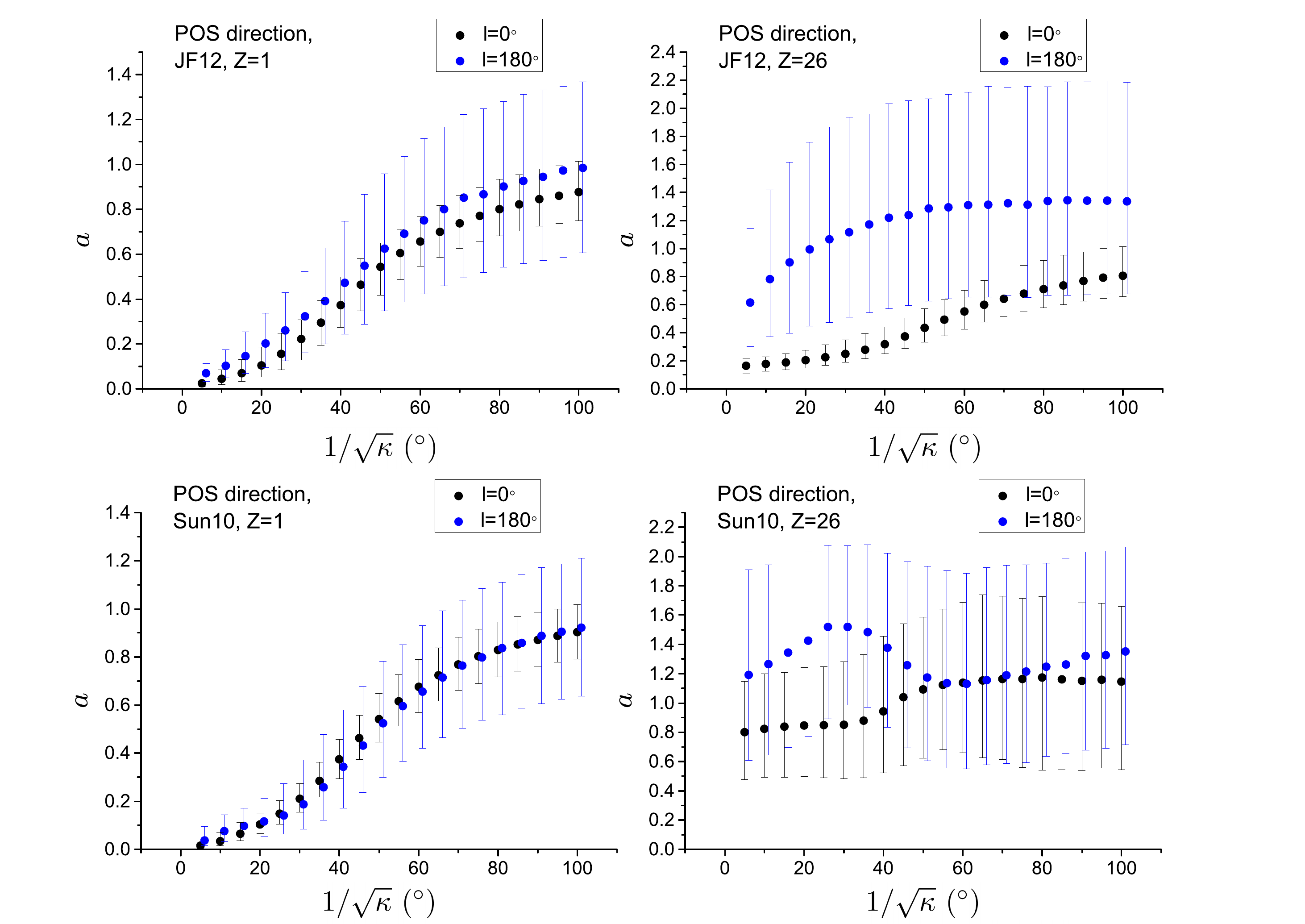}
\caption{As in figure \ref{fig:a_POSmagn}, but for the uncertainty $1/\sqrt{\kappa}$ of the POS direction.}
\label{fig:a_POSdir}
\end{figure}

First, we examine the dependence of the effectiveness coefficient $a$ on the uncertainty of the POS magnitude (figure \ref{fig:a_POSmagn}), the LOS magnitude (figure \ref{fig:a_LOSmagn}) and the POS direction (figure \ref{fig:a_POSdir}) of the magnetic field. Each of the three sources of uncertainty is examined independently. For example, when varying the uncertainty of the POS magnitude, the LOS magnitude and the POS direction are assumed to be known exactly. Details on what the uncertainty values on the horizontal axis of each figure represent can be found in \S  \ref{subsec:errors}.\footnote{We should stress that the quantity $1/\sqrt{\kappa}$ used in figure \ref{fig:a_POSdir} is only a useful measure of the uncertainty and does not in general represent the actual standard deviation of the distribution. Only for small uncertainties, where the von Mises distribution approaches a gaussian, does this quantity approximate the standard deviation. }

The values of $a$ were collected by using one specific realization of the GMF with 10000 iterations with different random errors. The points show the \textit{median} of the values of $a$ received for each value of uncertainty and the error bars show the 68\% confidence interval (i.e. the 68\% range of values around the median). This choice was made, instead of using the typically used mean and standard deviation, because the values of $\phi_{res}$ and $a$ were found to follow a quite skewed distribution, as can be seen from the asymmetry of the error bars. 

Our results are shown for two different values of the Galactic longitude and rigidity, and using two different GMF models. We assume particles with energy $E=60 \  {\rm EeV}$, charge $Z e$ and arrival direction of Galactic longitude $l$ and Galactic latitude $b=40 \degree$. 

Accuracies of $\sigma_{B}/B \leq 2$ are shown in figure \ref{fig:a_POSmagn} because, for larger uncertainties, blue points (arrival directions away from the Galactic center) generally exhibit prohibitively high effectiveness coefficients for any meaningful arrival direction corrections. The same range of accuracies is shown in figure \ref{fig:a_LOSmagn}, as the behavior of the effectiveness coefficient $a$ demonstrated for this range does not change abruptly for higher values of $\sigma_{B}/B$.

As expected, for small deflections (as is the case for 60 EeV protons), the sensitivity to the LOS magnetic field is very small, as the particle velocity stays approximately parallel to the LOS throughout its trajectory. In contrast, for larger deflections (as happens  for 60 EeV iron), a significant POS velocity component is present away from the Earth, making the correction sensitive to the LOS uncertainty. 

The figures show that a quite good correction is often feasible in the range of uncertainties that we have investigated. There are much better prospects of correction if the cosmic rays are protons, compared to iron nuclei of the same energy. There is also significant difference between arrival directions pointing towards the Galactic center ($l=0\degree$) and away from the Galactic center ($l=180\degree$). Cosmic rays with arrival directions closer to the Galactic center, where the deflection is larger, are good candidates for a better correction. Differences between the two GMF models can be very pronounced. For example, in the case of iron with $l=0\degree$, the JF12 model suggests that a good correction can be made, while the Sun10 model suggests that it is very difficult to make a correction. 

It is also apparent that some plots for iron have a more ``irregular'' form than for protons. This seems to happen in general for large deflections $(\gtrsim 90\degree)$ and especially when $a$ is also large. That is because $a$ and $\phi_{res}$ do not capture the actual details during the process of the trajectory reconstruction. For example, for some uncertainty value, the cosmic ray might statistically tend to go through a region of large magnetic field during the reconstruction, which will change its trajectory.

It should be noted that, while, as expected, the effectiveness coefficient (and so the residual angle) does decrease to zero as the uncertainty approaches zero, in the case of 60 EeV iron this decrease starts to happen at very low uncertainties. For example, using the JF12 model, an arrival direction with $l=180 \degree$ and $b=40 \degree$, and POS magnitude uncertainty of $\sigma_B/B \approx 10^{-3}$, the median of the effectiveness coefficient remains $a \approx 0.2$, while for $\sigma_B/B \approx 10^{-4}$, the median is $a \approx 0.01$.

\begin{figure}[hbtp] 
\centering
\includegraphics[width=0.93\textwidth]{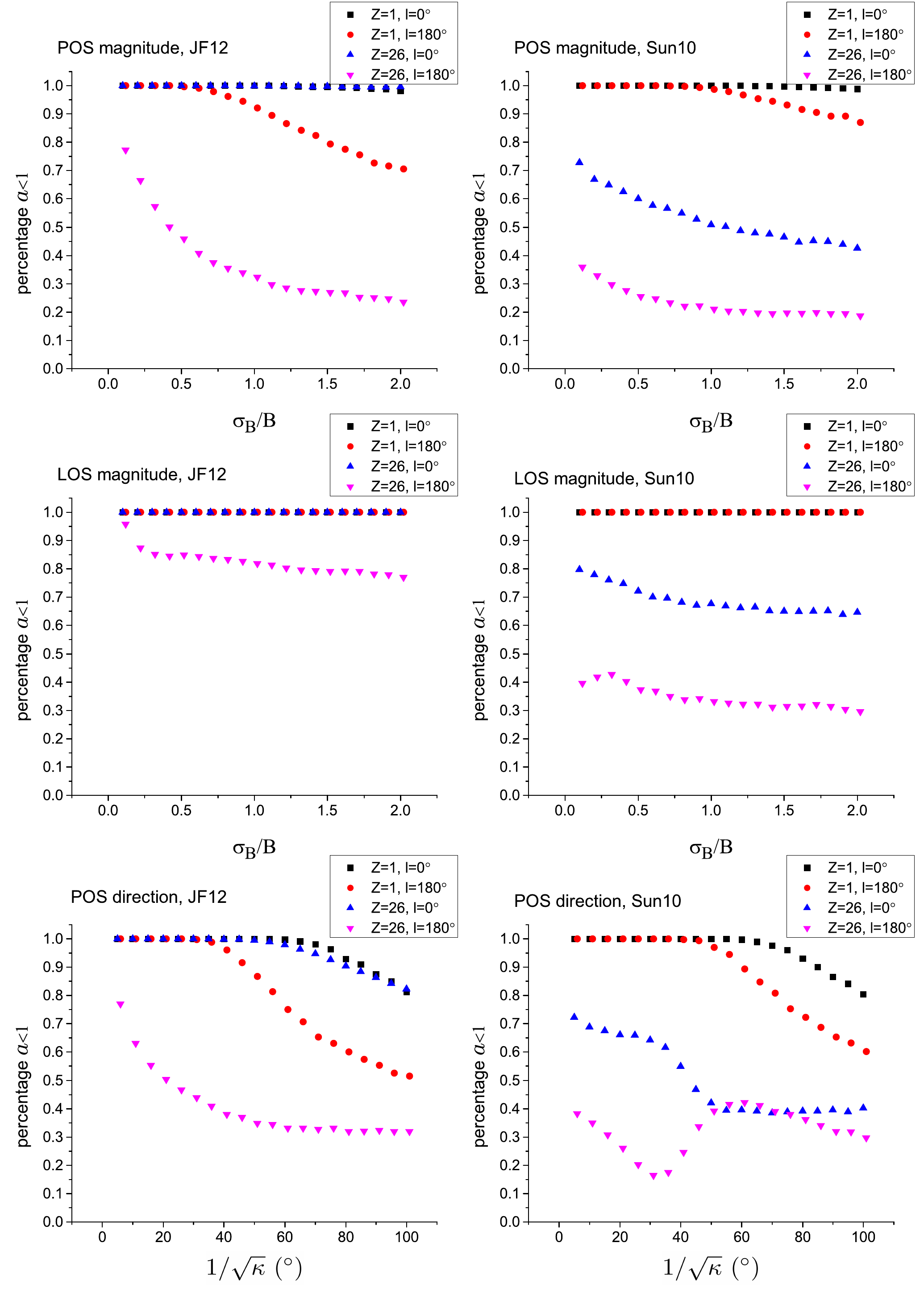}
\caption{Percentage of cases with effectiveness coefficient $a < 1$, for the same parameters and choices of uncertainties as those in figures \ref{fig:a_POSmagn}, \ref{fig:a_LOSmagn} and \ref{fig:a_POSdir}. }
\label{fig:p100}
\end{figure}

Some of the figures show that it is in some cases possible for $a$ to exceed the value of 1. This means that the residual angle becomes larger than the deflection. In that case, instead of being corrected, the deflection has been amplified. A useful question, then, is to ask what the probability of getting $a<1$ and thus managing to shrink the deflection is. This question is approximately answered in figure \ref{fig:p100}, where the percentage of the values of $a$ which satisfy $a<1$ is given for the same cases as before.

\begin{figure}[t]  
\includegraphics[trim= 30 270 30 30,clip,width=\textwidth]{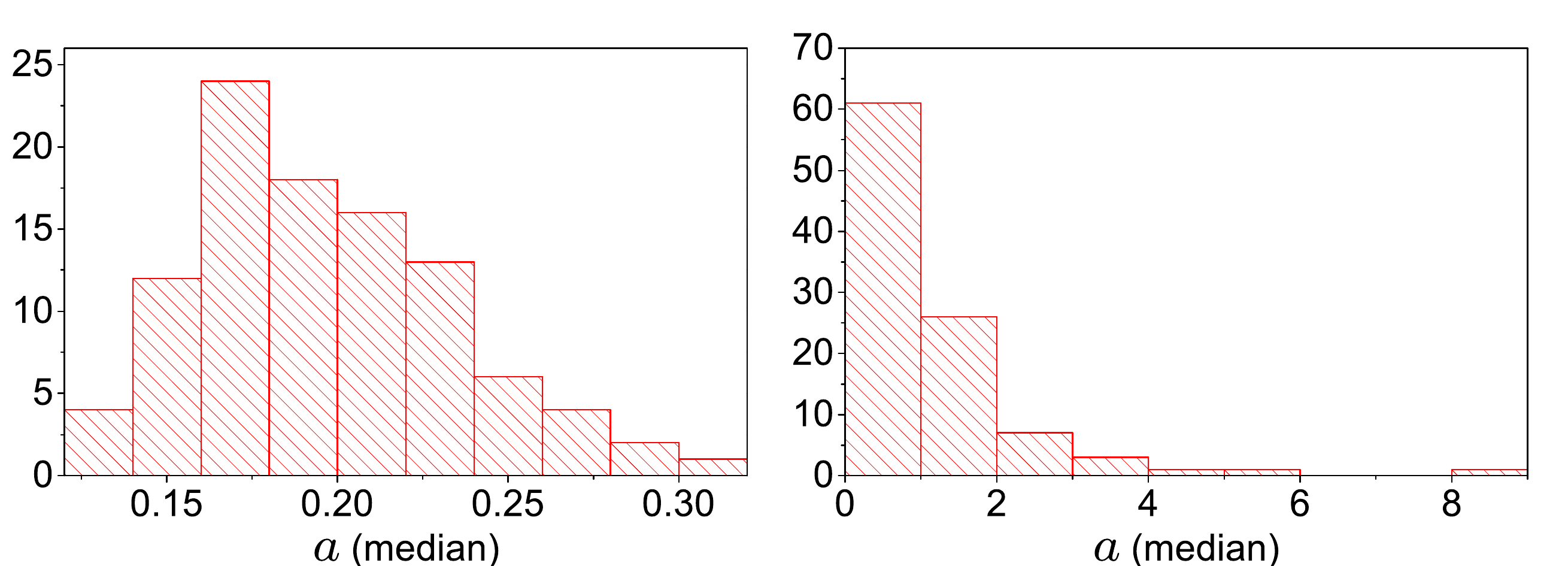}
\caption{Histograms showing the distribution of the median of the effectiveness coefficient $a$ for different random field realizations, assuming arrival directions with Galactic longitudes $l=0\degree$ (left panel) and $l=180\degree$ (right panel). See text for discussion of other parameters.}
\label{fig:a_histogram}
\end{figure}

The results are quite sensitive to the realization of the random magnetic field. Thus, figures \ref{fig:a_POSmagn}-\ref{fig:p100} are only meant to showcase some general features, while the actual values of $a$ shown can vary significantly between different realizations.

To give an example of how the random field can affect our results, we have repeated the above correction process for 100 realizations of the random field, each with 500 iterations with different random errors. In figure \ref{fig:a_histogram} we present histograms showing the distribution of the median of the effectiveness coefficient $a$ for two different cases. Both cases assume $60 \  \rm{EeV}$ protons arriving from a Galactic latitude $b=40\degree$ and an uncertainty of $\sigma_{B}/B=1$ only for the POS magnitude of the GMF. The JF12 model is used. In the first case (left panel) the arrival direction has a Galactic latitude $l=0\degree$ and in the second case (right panel) $l=180\degree$. These results can be compared with the top left panel of figure \ref{fig:a_POSmagn} for $\sigma_B/B=1$. These distributions can change depending on the rigidity, arrival direction, uncertainty of the magnetic field values and on how dominant the random field is throughout the cosmic ray trajectory compared to the regular field.

\subsection{Sky maps}

\begin{figure}[t]

\begin{subfigure}{0.5\textwidth}
\includegraphics[trim= 45 30 42 25, clip,width=\textwidth]{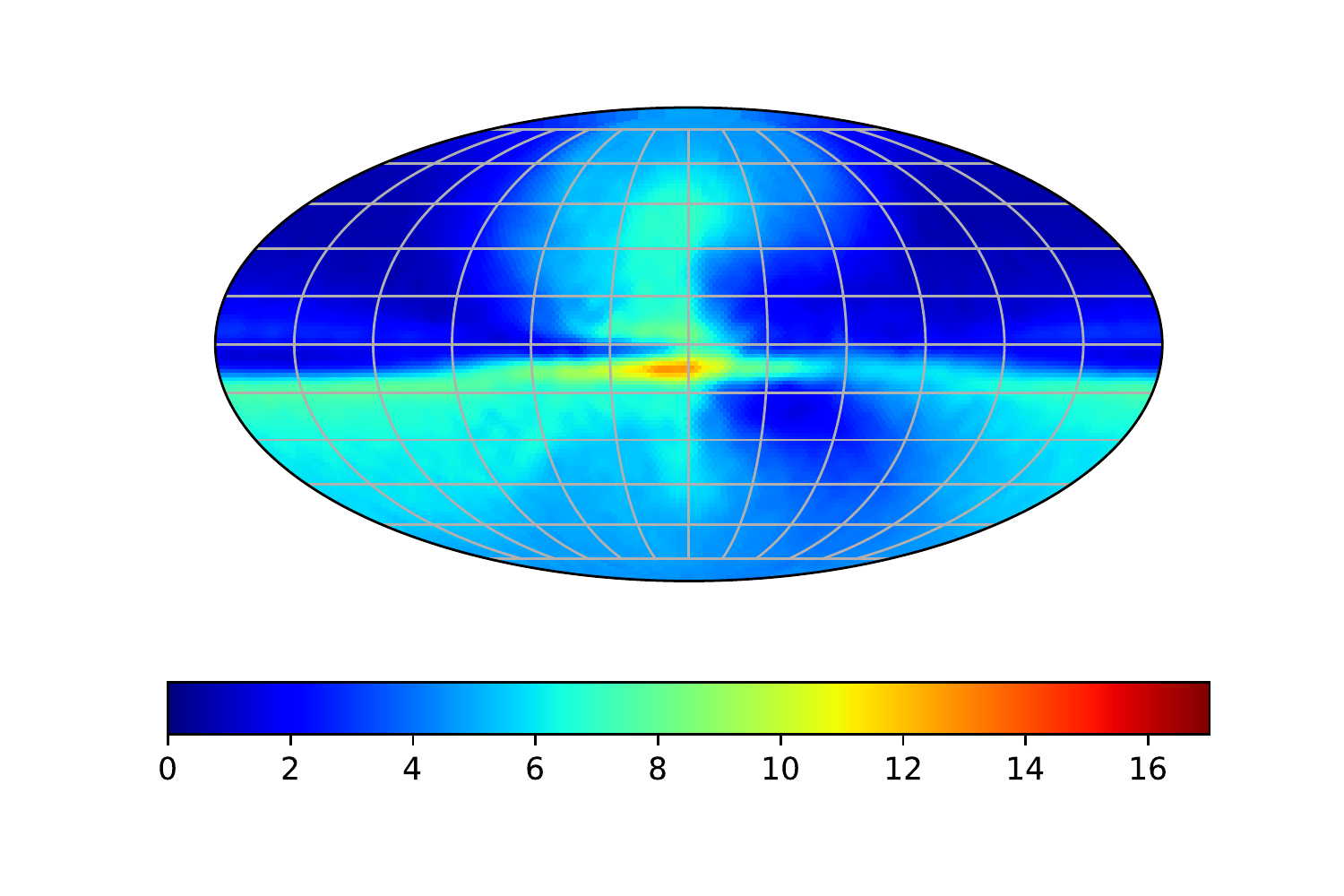}
\caption{Sky map of the median of the deflection for the JF12 model in degrees.}
\label{fig:JF12_deflection_map}
\end{subfigure}
\begin{subfigure}{0.5\textwidth}
\includegraphics[trim= 45 30 42 25, clip,width=\textwidth]{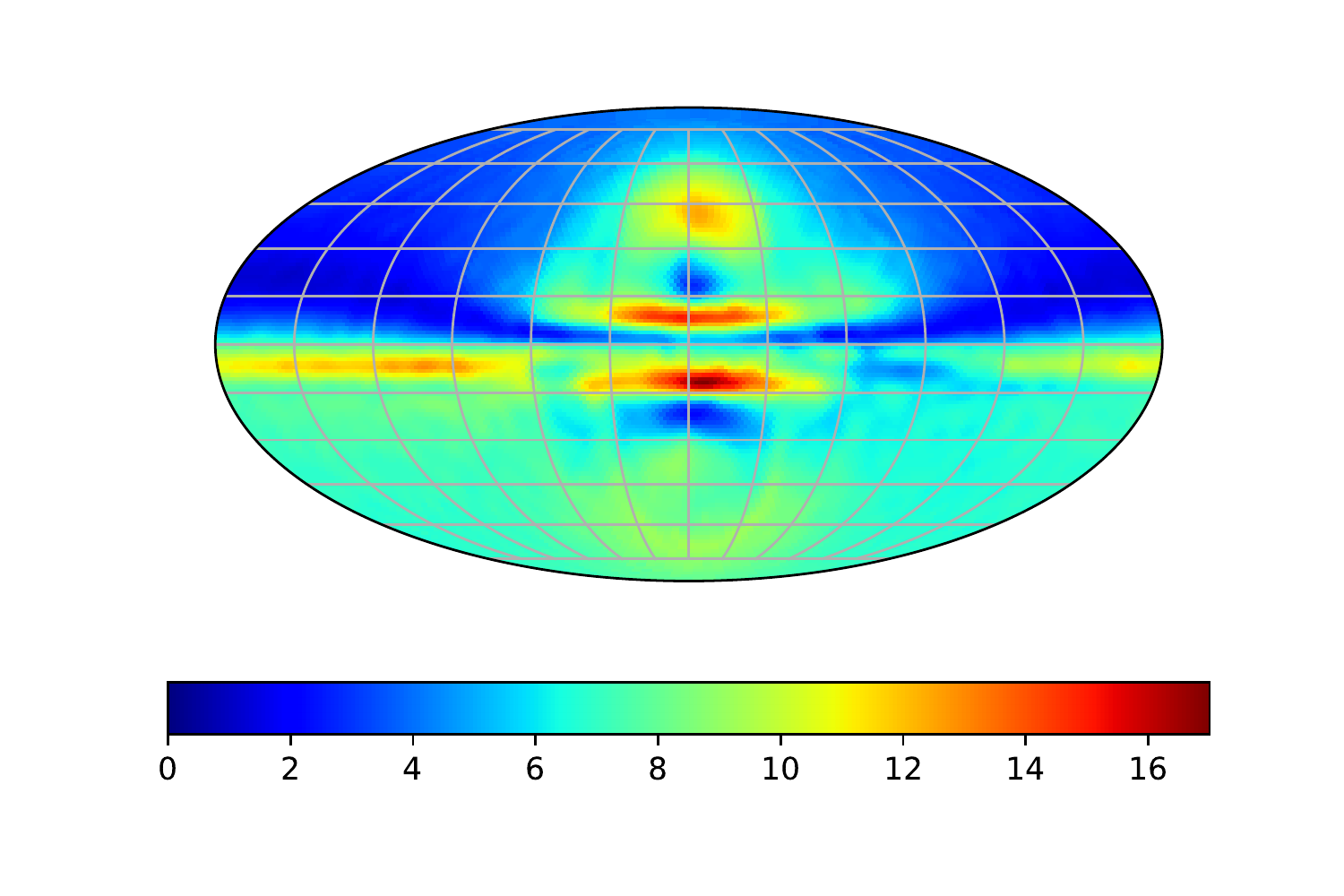}
\caption{Sky map of the median of the deflection for the Sun10 model in degrees.}
\label{fig:Sun10_deflection_map}
\end{subfigure}

\begin{subfigure} {0.5\textwidth}
\includegraphics[trim= 45 30 42 25, clip,width=\textwidth]{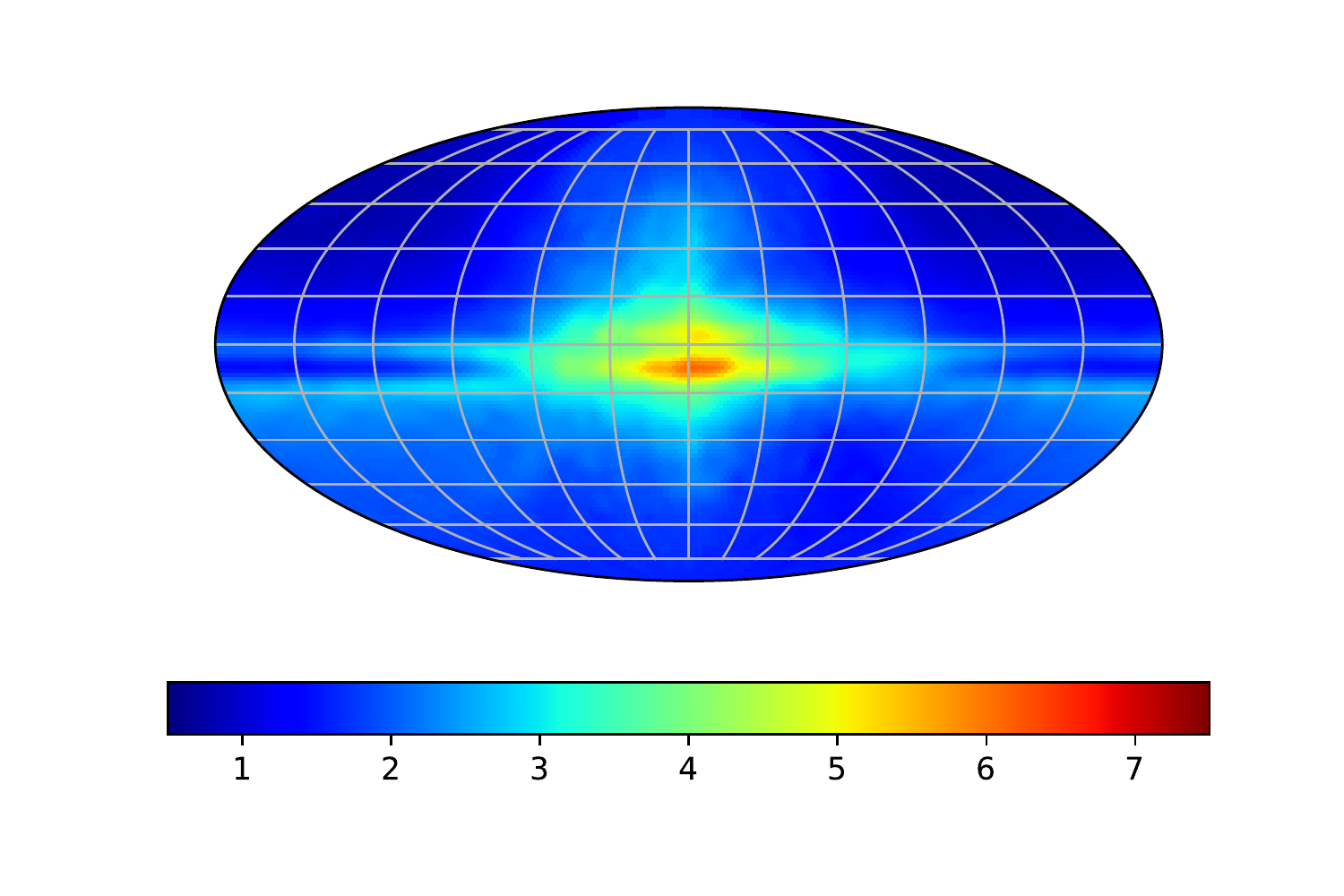}
\caption{Sky map of the median of the residual angle for the JF12 model in degrees.}
\label{fig:JF12_phires_map}
\end{subfigure}
\begin{subfigure} {0.5\textwidth}
\includegraphics[trim= 45 30 42 25, clip,width=\textwidth]{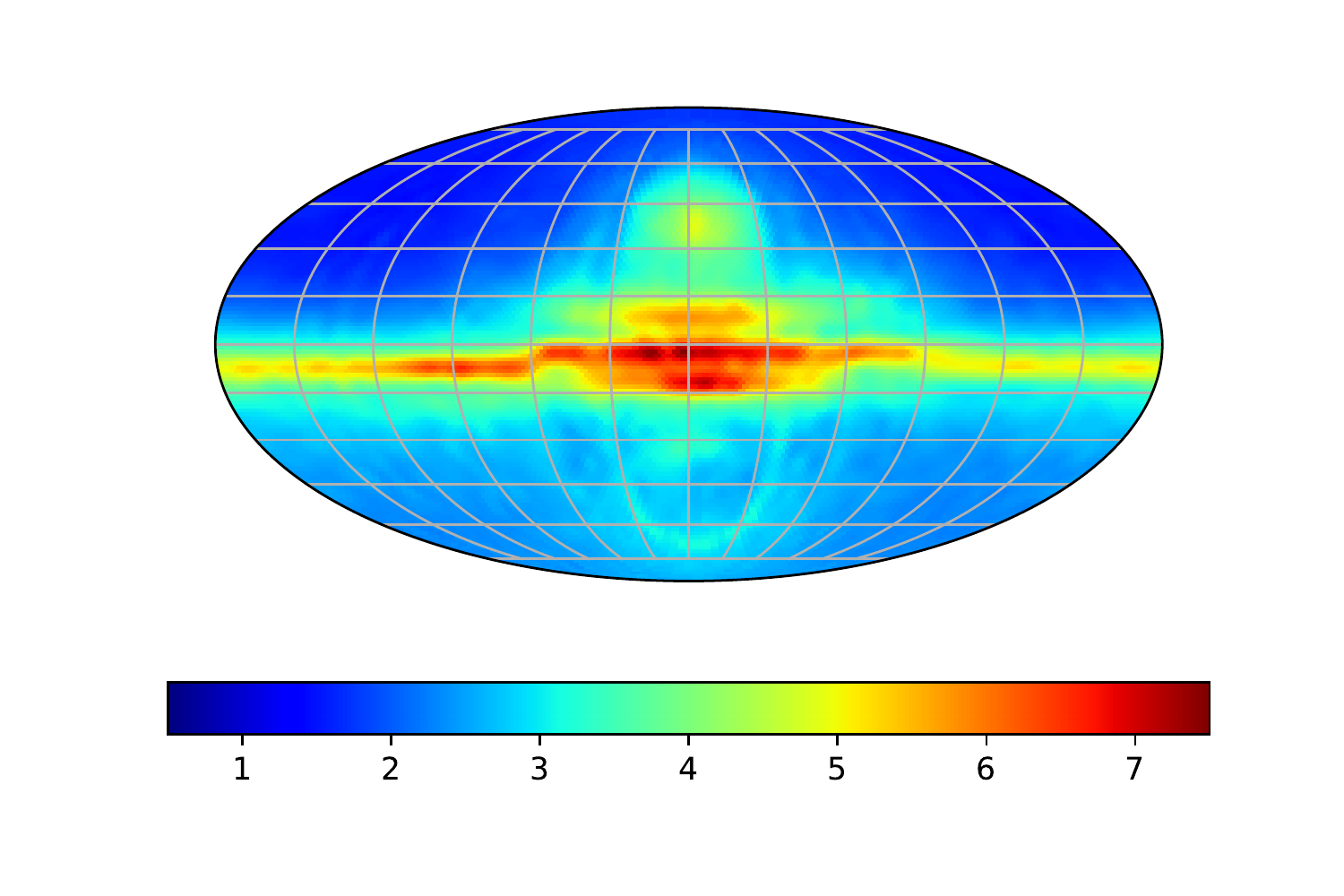}
\caption{Sky map of the median of the residual angle for the Sun10 model in degrees.}
\label{fig:Sun10_phires_map}
\end{subfigure}

\begin{subfigure} {0.5\textwidth}
\includegraphics[trim= 45 30 42 25, clip,width=\textwidth]{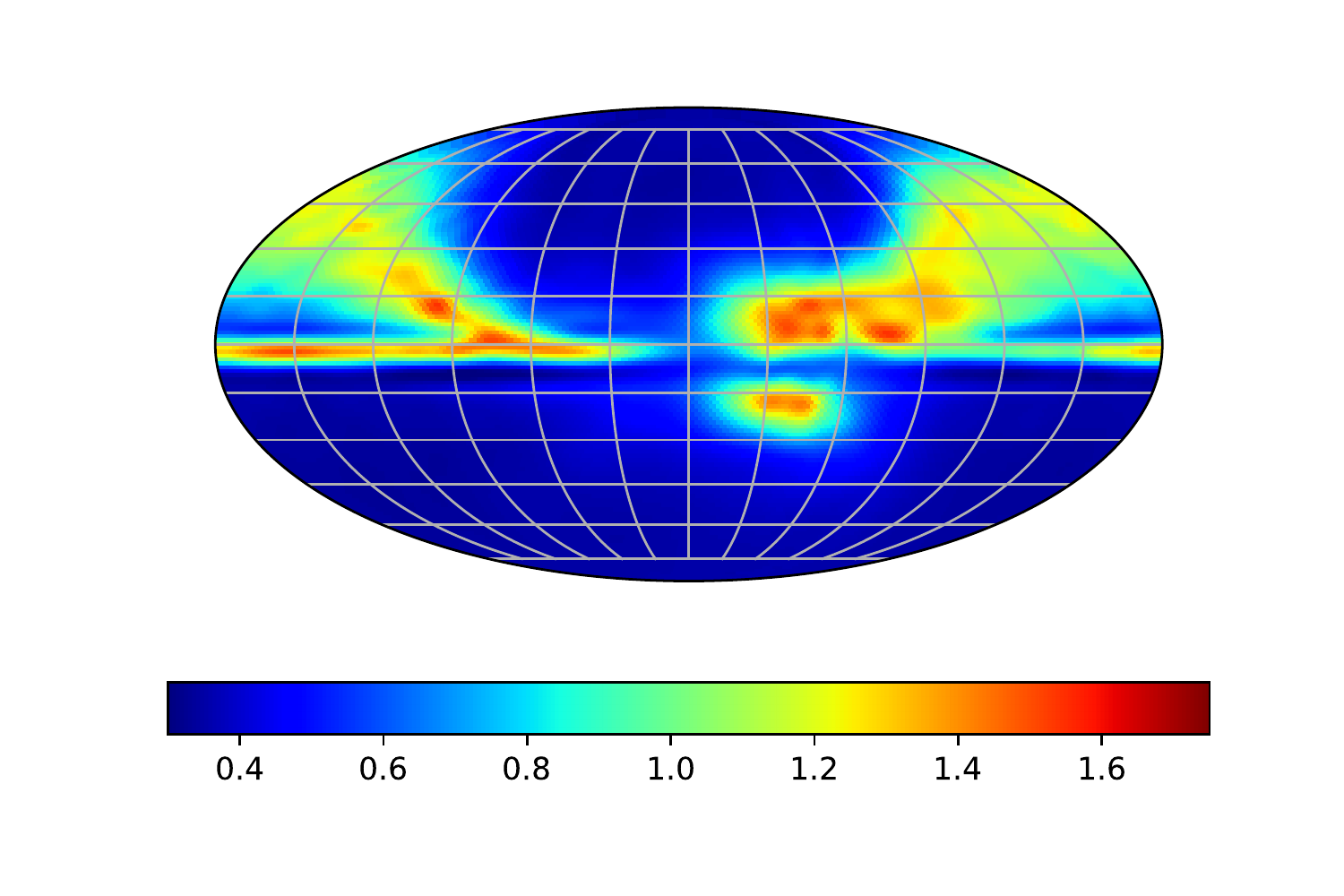}
\caption{Sky map of the median of the effectiveness coefficient for the JF12 model.}
\label{fig:JF12_alpha_map}
\end{subfigure}
\begin{subfigure} {0.5\textwidth}
\includegraphics[trim= 45 30 42 25, clip,width=\textwidth]{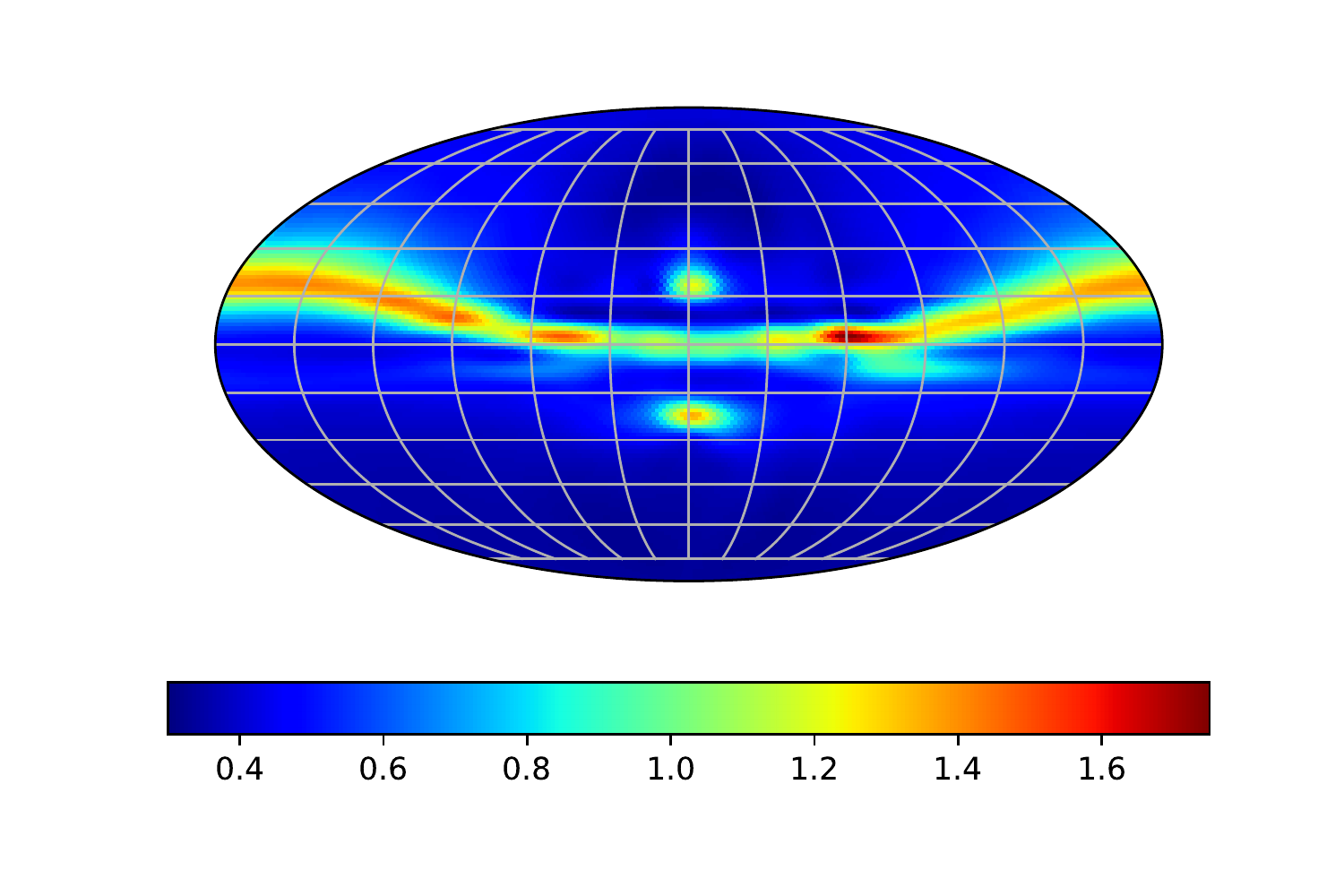}
\caption{Sky map of the median of the effectiveness coefficient for the Sun10 model.}
\label{fig:Sun10_alpha_map}
\end{subfigure}

\caption{Sky map in Mollweide projection, using Galactic coordinates, of the median of (a),(b) the deflection, (c),(d) the residual angle and (e),(f) the effectiveness coefficient for 60 EeV protons, using the JF12 and Sun10 models respectively. The uncertainty values used are mentioned in the text. Galactic longitude increases to the left.}
\label{fig:sky_map}
\end{figure}

To showcase how the effectiveness of the correction depends on the arrival direction of the cosmic ray, we have constructed sky maps in Mollweide projection of the median of the deflection, the residual angle and the effectiveness coefficient. These are shown in figure \ref{fig:sky_map}. The results are for 60 EeV protons and uncertainties of $\sigma_B/B=1$ for the POS and LOS magnitudes and $1/\sqrt{\kappa}=40 \degree$ for the POS direction, using both the JF12 and Sun10 models of the GMF.
\footnote{For the JF12 model, the parameter $r_s$, which describes the toroidal halo field for the southern Galactic hemisphere (see appendix \ref{sec:GMF}), remains unspecified as only a lower limit is given to be 16.7 kpc. In our simulation, we have considered its value to be equal to that lower limit. While the previous results are not affected by the choice of $r_s$, because they concern the northern Galactic hemisphere, these results are somewhat affected. After constructing the same sky maps using the highest possible value of 20 kpc, it was found that the residual angle and deflection angle were slightly affected in the southern hemisphere, but the effectiveness coefficient remained relatively unchanged, at least for the choice of uncertainties and rigidity tested.}
To create these maps, we performed simulations for 2520 arrival directions in total, covering the whole sky, with the Galactic longitude ranging from $-180 \degree$ to $175 \degree$ with a step of $5\degree$, and the Galactic latitude from $-85 \degree$ to $+85 \degree$, again with a $5\degree$ step.

In this case, the correction process was repeated for 100 realizations of the magnetic field and 50 iterations with different random errors for each realization. Values for the deflection, the residual angle and the effectiveness coefficient were collected during each iteration and the median of all collected values was calculated. Thus, here the median is over both the different random field realizations and the iterations with different random errors (of course, the deflection changes only with the different realizations). This way, we can get a sense of the ``general'' behaviour of each GMF model, independent of the specific realization of the random field.

Several features can be observed from these sky maps. First, it is apparent that there are significant differences between the two models, as seen most dramatically in the deflection sky maps. The Sun10 model predicts in general larger deflections than the JF12 model, while the location of regions of large deflection also differs. Next, the residual angle tends to be larger for arrival directions near the Galactic center, which is expected, as the deflection is also larger. Also, looking at the sky maps of the effectiveness coefficient $a$, we can see that, to a large extent, their form can be predicted from the respective deflection maps. Arrival directions with larger deflection have lower $a$.

\section{Discussion}

We have found that there are important differences between the two GMF models used and also that the random component can influence significantly the effectiveness of the correction. This showcases the fact that our knowledge of the GMF is still quite limited, at least for the purposes of reconstructing the trajectory of UHECRs, and highlights the importance of making actual 3D measurements of the GMF. Our most significant differences between different models arise in the southern Galactic hemisphere, in agreement with the results of \cite{uncertainty}, who have recently studied the uncertainties involved in the parametrization of the GMF. After an experiment has been carried out to make such measurements, a similar analysis to that presented in this paper can be performed using the specific measurements and error distribution of that experiment, in order to estimate more accurately the uncertainty of any backtracing.

It should be noted that, as reported in \cite{Deflections}, which provides a detailed study of the magnetic deflections of UHECRs using the JF12 model, the deflections are found to be sensitive to the coherence length of the random component of the GMF. Thus, the coherence length that we have assumed in our simulations might have influenced our results for the effectiveness of the correction as well. It is also possible that the coherence length presents large variations throughout the Galaxy, while we have assumed it to be the same everywhere in the Galaxy. 

It is important to stress that our work has assumed random errors within each grid cube, which are only realizable in a bona-fide tomographic experiment rather than the currently-available parameter-fitting modeling. As such, we strongly caution the reader that our results cannot be regarded as predictive of the effectiveness of deflection corrections using current GMF models. 

For completeness, we have assumed that tomographic measurements of the GMF are available everywhere in the sky. However, any specific experiment might yield measurements for limited parts of the sky only. In addition, we have used a spatial resolution of 100 pc as a reasonable choice for the coherence length of the Galactic magnetic field, as our purpose here is to examine the effect of measurement uncertainties alone, assuming that the sampling is dense enough to provide an adequate/complete set of measurements. Additional uncertainties in the de-propagation may, thus, be introduced by incomplete sampling along the cosmic ray trajectory.

Finally, we found that the effectiveness of the correction is likely to be high for light cosmic ray composition. Therefore, given relatively accurate measurements of the GMF, the process of correction can be used as a composition probe. If no increased clustering is found assuming proton or light nuclei composition, where a source and good correction can be reasonably expected, then this result can be interpreted as evidence for a heavy composition. 

\section{Summary and Conclusions}

In our present work, we have focused on the possibility of correcting the magnetic deflection of UHECRs using 3D measurements of the GMF. We have attempted to examine how the effectiveness of such a correction depends on the uncertainty of our measurements and how it is influenced by the rigidity and arrival direction of the cosmic ray, assuming that these are perfectly known. To that end, we have constructed a numerical code, which simulates attempts to make such corrections using hypothetical measurements of the GMF, based on two recently updated GMF models, with varying uncertainty for the POS magnitude, the LOS magnitude and the POS direction of the magnetic field. 

Our results highlight the conditions under which an effective correction is achievable. A better correction can be achieved in general if the UHECRs are of high rigidity (ideally if they are high-energy protons) and if they have arrival directions that imply a larger deflection, which can be more easily corrected. In the case of ultra-high-energy protons, knowledge of the magnitude of the LOS component of the magnetic field is unnecessary, while in the case of iron nuclei, it becomes of similar importance to the POS component.

The significant differences in predictions using different GMF models underlines the sensitivity of our results to the actual GMF geometry. This however will be much better constrained should a 3D tomographic GMF mapping become available. 

\acknowledgments

We wish to thank Kostas Tassis and Aris Tritsis for valuable discussions that helped to improve the present work.

\appendix

\section{Galactic Magnetic Field Models}
\label{sec:GMF} 

Here we briefly present the GMF models that we have used in our code. These are the Jansson \& Farrar model (JF12) \cite{JF12_ord,JF12_rand} and the Sun \& Reich model (Sun10) \cite{Sun8,Sun10}. Both of these models have recently been updated by the Planck Collaboration \cite{Planck}, and these updates, which will also be mentioned, have been implemented into our code. 

The parameters in the formulas below that are not explained are free parameters of the models, and the values that we have used for each of them (based on the original papers and the Planck update) can be found in the respective tables. 

\subsection{JF12}

The JF12 magnetic field consists of a large-scale regular component, a random component and a ``striated'' component, often mentioned in other models as the ``ordered random'' component. Our choice of parameters is based on the updated model called ``Jansson12b'' in the Planck update. 

\subsubsection{Regular Component}

The regular field for this model consists of a disk component, a toroidal halo component and an out-of-plane component, also referred to as the ``X-field'' component. The parameter values used in our code for the regular component can be found in table \ref{tab:JF12 regular}. This table also shows which of these parameter values have been updated by the Planck Collaboration and which have not. 

The coordinates used are Cartesian $(x,y,z)$, as well as cylindrical $(r,\phi ,z)$, where the Galactic center is placed at the origin, the Galactic plane lies on the $x-y$ plane, and the Sun is placed at $x=-8.5 \ {\rm kpc}$ and $y=0$. The Galactic North is towards the positive z-direction.  

\bigskip

The disk field is defined for $3 \ {\rm kpc} \leq r \leq 20 \ {\rm kpc}$. For $3 \ {\rm kpc} \leq r \leq 5 \ {\rm kpc}$, there is a ``molecular ring'', which has a purely azimuthal field $\bm{b} = b_{ring} \hat{\bm{\phi}}$. For larger $r$, eight logarithmic spiral regions are defined, with boundaries obeying the equation 

\begin{equation}
r=r_{-x}\exp \left[(\phi-\pi) \tan i \right] ,
\end{equation}
where $i=11.5 \deg$ is the opening angle of the spirals and $r_{-x}$ is the radius where each spiral crosses the negative $x$-axis.\footnote{Note typo in the original paper. Also, we have changed the origin of the angle $\phi$ so that $\phi = \pi$ at the negative x-axis.} We take $r_{-x}= \lbrace 5.1, 6.3, 7.1, 8.3, 9.8, 11.4, 12.7, 15.5 \rbrace {\rm kpc} $. The disk field is then given by the formula

\begin{equation}
b_{disk}= (1-L(z,h_{disk},w_{disk})) \cdot  b_i \cdot 5 \ {\rm kpc} / r ,
\end{equation}
where $b_i$ is the field strength of the i-th spiral region at $r=5\ {\rm kpc}$, and

\begin{equation}
L(z,h,w)=(1+\exp \left( -2(\vert z \vert - h)/w\right))^{-1}
\end{equation}
is a logistic function used to describe the transition from the disk field to the halo field. The field direction is $\hat{\bm{b}}_{disk}= \sin i \  \ru + \cos i \  \phiu $.

\bigskip

The toroidal halo field, as its name implies, has a purely azimuthal component. For $z>0$ (north half of the halo), the field is 

\begin{equation}
B_{\phi}^{tor} = \exp \left(-\vert z \vert /z_0 \right) L(z,h_{disk},w_{disk}) B_n (1-L(r,r_n,w_h)) ,
\end{equation}
while for $z<0$ the parameters $B_n$ and $r_n$ are substituted by the different parameters $B_s$ and $r_s$ respectively.\footnote{As can be seen in table \ref{tab:JF12 regular}, the parameter $r_s$ is not specified, but only a lower limit is given. In our code we have chosen $r_s$ to be equal to this lower limit.}
Thus, the field strength and radial extent of the halo field are different for the north and south halves of the halo. 

\bigskip

Finally, the ``X-field'' component is axisymmetric and purely poloidal (i.e. it does not have an azimuthal component). Thus the field changes only in the $(r,z)$ plane, remaining the same for any angle $\phi$. The field lines are straight lines, which run from the southern direction towards the z-axis with a certain ``elevation angle'' $\Theta_X$ with the $z=0$ plane, and, when they cross the $z=0$ plane, they run away from the $z$-axis and towards the northern direction with the same angle $\Theta_X$. The radius at which a field line crosses the $z=0$ plane is defined as $r_p$. To every point $(r,z)$ corresponds one field line, and therefore one radius $r_p$. If $r_p$ is larger than a certain radius $r_X^c$, the field lines are taken to have a constant elevation angle $\Theta_X^0$. If $r_p < r_X^c$, then the elevation angle increases as $r_p$ approaches zero, and reaches the value $\Theta_X = 90 \degree $ at $r_p=0$.  

The strength of the magnetic field at the $z=0$ plane is defined as 

\begin{equation}
b_X (r_p)=B_X \exp (-r_p/r_X) , 
\end{equation}
where $B_X$ and $r_X$ are free parameters of the model. With the above definitions and geometry, the requirement $\nabla \cdot \bm{B} = 0$ gives the following formulas for the magnetic field.

When 

\begin{equation}
r_p \geq r_X^c \Leftrightarrow r \geq r_X^c + \frac{z}{\tan{\Theta_X^0}},
\end{equation}
the magnetic field strength is

\begin{equation}
b_X(r)= b_X(r_p) \cdot r_p/r ,
\end{equation}
with

\begin{equation}
r_p=r-\frac{\mid z \mid}{\tan \Theta_X^0} .
\end{equation}

When $r_p < r_X^c$, the magnetic field strength is

\begin{equation}
b_X(r)=b_X(r_p) \left(r_p/r\right) ^2,
\end{equation}
with

\begin{equation}
r_p=\frac{r \  r_X^c}{r_X^c + \mid z \mid / \tan \Theta_X^0 } ,
\end{equation}
and the elevation angle changes as 

\begin{equation}
\Theta_X = \tan ^{-1} \left(\frac{\mid z \mid }{r -r_p} \right) .
\end{equation}

\begin{table} [h!]
\begin{center}
\begin{tabular}{|c|c|c|}

\hline 

Parameter & Value & Updated? \\
\hline
\hline 

$b_1$ & 0.1 $\mu$G & No \\
$b_2$ & 3.0 $\mu$G & No \\
$b_3$ & -0.9 $\mu$G & No \\
$b_4$ & -0.8 $\mu$G & No \\
$b_5$ & -2.0 $\mu$G & No \\
$b_6$ & -3.5 $\mu$G & Yes \\
$b_7$ & 0.0 $\mu$G & No \\
$b_8$ & 2.7 $\mu$G & No \\
$b_{ring}$ & 0.1 $\mu$G & No \\
$h_{disk}$ & 0.4 kpc & No \\
$w_{disk}$ & 0.27 kpc & No \\

\hline 

$B_n$ & 1.4 $\mu$G & No \\
$B_s$ & -1.1 $\mu$G & No \\
$r_n$ & 9.22 kpc & No \\
$r_s$ & $>$ 16.7 kpc & No \\
$w_h$ & 0.2 kpc & No \\
$z_0$ & 5.3 kpc & No \\

\hline 

$B_X$ & 1.8 $\mu$G & Yes \\
$\Theta_X^0$ & 49$\degree$ & No \\
$r_X^c$ & 4.8 kpc & No \\
$r_X$ & 2.9 kpc & No \\

\hline

\end{tabular}
\caption{Parameter values used for the JF12 regular field.}
\label{tab:JF12 regular}
\end{center}
\end{table}

\subsubsection{Random Component}

The random component is taken to be isotropic and its root mean square (RMS) value has different contributions from the disk and the halo. Parameter values can be found in table \ref{tab:JF12 random}.

The disk random field is given by $$B_{disk}^{rand} = f(r) \cdot e^{-z^2/2 (z_0^{disk})^2}$$ where 

\begin{equation}
   f(r) = \left\{\begin{array}{lr}
        b_{int}, & \text{for }r < 5 \ {\rm kpc} \\
        b_i \cdot \frac{5 \ {\rm kpc}}{r}, & \text{for } r \geq 5 \ {\rm kpc}
        \end{array}\right\} 
\end{equation}

The value of the parameter $b_i$ is different in each spiral region defined as in the regular field (note that this parameter is different from the regular component parameter of the same name; we understand that this choice of notation might be somewhat confusing, but we have decided to retain the notation of the original papers).

The halo field is 

\begin{equation}
B_{halo}^{rand}= B_0 e^{-r/r_0} e^{-z^2/2 z_0^2}. 
\end{equation}

The total random field strength is then taken to have the RMS value 

\begin{equation}
B_{rand}= \sqrt{(B_{disk}^{rand})^2 + (B_{halo}^{rand})^2} .
\end{equation}

To implement this field in our code, we use three random gaussian variables with a mean of 0 and a standard deviation of $B_{rand}/ \sqrt{3}$ , which represent the x, y and z components of the magnetic field. This choice satisfies the requirements that the field is isotropic and that the RMS value for the magnetic field strength is $B_{rand}$.

\begin{table} [h!]
\begin{center}
\begin{tabular}{|c|c|c|}

\hline 

Parameter & Value & Updated? \\
\hline
\hline 

$\langle B_{iso}^2 \rangle $ & 5 $\mu$G & Yes \\

\hline

$b_{even}$ & 0.8 $\langle B_{iso}^2 \rangle $ & Yes \\

$b_{odd}$ & 0.4 $\langle B_{iso}^2 \rangle $ & Yes \\

$b_{int}$ & 0.5 $\langle B_{iso}^2 \rangle $ & Yes \\

$z_0^{disk}$ & 0.61 kpc & No \\

\hline 

$B_0$ & 0.94 $\langle B_{iso}^2 \rangle $ & Yes \\

$r_0$ & 10.97 kpc & No \\
$z_0$ & 2.84 kpc & No \\

\hline

\end{tabular}
\caption{Parameter values used for the JF12 random field.}
\label{tab:JF12 random}
\end{center}
\end{table}

\subsubsection{Striated Component}

The JF12 model also includes a ``striated'' component, often called in other models the ``ordered random'' component. A striated field is thought to be produced when an isotropic random or perhaps a coherent field experiences stress or shear. It is assumed to have a preferred large-scale direction parallel to the regular field, but experiences sign reversals on a small scale, thus its average value is zero. 

We have neglected this component in our code. A discussion of its effects on UHECR propagation can be found in \cite{Deflections_Centaurus} and a specific way to implement it is mentioned in \cite{Keivani_diss}.

\subsection{Sun10}

The paper presenting the Sun10 model actually contains three distinct regular field models, as well as a simple treatment for the random component. We focus on the model that has been updated by the Planck Collaboration, which is the ASS+RING model. ASS stands for axi-symmetric spiral and RING refers to its structure, as it uses concentric rings in which the sign of the field is reversed. Throughout this paper, the updated ASS+RING model is referred to simply as Sun10. In the Planck update, a more detailed random component, as well as an ordered random field (the equivalent of the striated field in the JF12 model) have been added to this model. As in the JF12 model, we have neglected the ordered random component. 

\subsubsection{Regular Component}

The regular component of the Sun10 model is significantly simpler than that of the JF12 model. It includes a disk component $\bm{B}^D$ and a halo component $\bm{B}^H$. Parameter values can be found in table \ref{tab:Sun10 regular}.

The disk component is written in cylindrical coordinates $(R,\phi,z)$ with the Galactic centre at the origin as: 

\begin{align}
B_R^D &= D_1(R,\phi,z) D_2(R,\phi,z) \sin p \\
B_{\phi}^D &=-D_1(R,\phi,z) D_2(R,\phi,z) \cos p \\
B_z^D &=0
\end{align}
where 

\begin{equation}
   D_1(R,z) = \left\{\begin{array}{lr}
        B_0 \exp \left(-\frac{R-R_{\bigodot}}{R_0} - \frac{\mid z \mid}{z_0} \right), & \text{for } R>R_c\\
        B_c, & \text{for } R \leq R_c
        \end{array}\right\} 
\end{equation}
with $R_{\bigodot}=8.5 \ {\rm kpc}$ being the Galactic radius of the Sun, and 

\begin{equation}
    D_2(R) = \left\{\begin{array}{lr}
        +1, & \text{for } R>7.5 \ {\rm kpc} \\
       -1, & \text{for } 6 \ {\rm kpc} < R \leq 7.5 \ {\rm kpc}\\
        +1, & \text{for } 5 \ {\rm kpc} < R \leq 6 \ {\rm kpc} \\
        -1, & \text{for } R \leq 5 \ {\rm kpc}
        \end{array}\right\}.
\end{equation}

The halo field is purely azimuthal and is written as: 

\begin{equation}
B_{\phi}^H (R,z) = \text{sign} (z) \  B_0^H \frac{1}{1+ \left( \frac{\mid z \mid - z_0^H}{z_1^H} \right) ^2} \frac{R}{R_0^H} \exp \left( - \frac{R-R_0^H}{R_0^H} \right).
\end{equation}

\begin{table} [h!]
\begin{center}
\begin{tabular}{|c|c|c|}

\hline 

Parameter & Value & Updated? \\
\hline
\hline 

$R_0$ & 10 kpc & No \\
$z_0$ & 1 kpc & No \\
$R_c$ & 5 kpc & No \\
$B_0$ & 2 $\mu$G & No \\
$B_c$ & 0.5 $\mu$G & Yes \\

\hline

$z_0^H$ & 1.5 kpc & No \\
$z_1^H$ & 0.2 kpc, for $|z|<z_0^H$ & No \\
	& 0.4 kpc, otherwise & No \\
$B_0^H$ & 10 $\mu$G & No \\
$R_0^H$ & 4 kpc & No \\

\hline

\end{tabular}
\caption{Parameter values used for the Sun10 regular field.}
\label{tab:Sun10 regular}
\end{center}
\end{table}

\subsubsection{Random Component}

The random component in the updated Sun10 model has an RMS strength that can be written as: 

\begin{equation}
B_{RMS} (R,z) = \langle B_{iso}^2 \rangle ^{1/2} f(R) g(z)
\end{equation}
where 

\begin{equation}
f(R)= \exp \left(- \frac{R-R_{\bigodot}}{r_0^{ran}} \right) ,
\end{equation}
and 

\begin{equation}
g(z)= (1-f_{disk}^{ran}) {\rm sech} ^2 \left( \frac{z}{h_{halo}^{ran}} \right) 
+ f_{disk}^{ran} {\rm sech} ^2 \left( \frac{z}{h_{disk}^{ran}} \right) .
\end{equation}

Again, to model this isotropic random field in our code, we use three random gaussian variables with a mean of 0 and a standard deviation of $B_{RMS}/\sqrt{3}$. Parameter values can be found in table \ref{tab:Sun10 random}.

\begin{table} [h!]
\begin{center}
\begin{tabular}{|c|c|c|}

\hline 

Parameter & Value & Updated? \\
\hline
\hline 

$\langle B^2_{iso} \rangle$ & 4.8 $\mu$G & Yes \\

\hline

$r_0^{ran}$ & 30 kpc & Yes \\
$f_{disk}^{ran}$ & 0.5 & Yes \\
$h_{halo}^{ran}$ & 3 kpc & Yes \\
$h_{disk}^{ran}$ & 1 kpc & Yes \\

\hline

\end{tabular}
\caption{Parameter values used for the Sun10 random field.}
\label{tab:Sun10 random}
\end{center}
\end{table}

\section{The von Mises distribution}
\label{sec:vonMises}

To simulate a random angular variable, one needs to use a circular distribution.
The generalization of the normal distribution to circular variables is called the wrapped normal distribution. In our simulation, we have chosen to use the von Mises distribution, which involves simpler computations and closely approximates the wrapped normal distribution. 

A random angular variable $\Theta$ obeying the von Mises distribution $VM(\mu,\kappa)$, has a probability density function of the form

\begin{equation}
f(\theta) = \frac{1}{2 \pi I_0(\kappa)} e^{\kappa \cos(\theta - \mu)} ,
\end{equation}
where $I_0$ is the modified Bessel function of the first kind and order 0, $\mu$ is the mean of the distribution, and $\kappa$ is the ``concentration parameter'' \cite{Dir_stat}. For large $\kappa$ the von Mises distribution approaches a normal distribution with standard deviation $1/\sqrt{\kappa}$. 

In our code, the von Mises distribution is simulated using the method of \cite{von_Mises}.


\begin{thebibliography}{99}

\bibitem{Aloisio}
R. Aloisio, \emph{The Physics of UHECRs: Spectra, Composition
and the Transition Galactic-Extragalactic}, (2017)
[arXiv:1704.07110]

\bibitem{TA}
R.~U. Abbasi \textit{et al.}, \emph{The energy spectrum of cosmic rays above 10$^{17.2}$ eV measured by the fluorescence detectors of the Telescope Array experiment in seven years}, \emph{Astropart. Phys.} {\bf 80} (2016) 131.

\bibitem{Han}
J. L. Han, \emph{Observing Interstellar and Intergalactic Magnetic Fields}, \emph{Annu. Rev. Astron. Astrophys.} {\bf 55} (2017) 111.

\bibitem{Planck}
Planck Collaboration, \emph{Planck intermediate results. XLII. Large-scale Galactic magnetic fields}, \emph{Astron. Astrophys.} {\bf 596} (2016) A103. [arXiv:1601.00546]

\bibitem{JF12_ord}
R. Jansson and G. Farrar, \emph{A New Model of the Galactic Magnetic Field}, \emph{Astrophys. J.} {\bf 757} (2012) 14. [arXiv:1204.3662]

\bibitem{JF12_rand}
R. Jansson and G. Farrar, \emph{The Galactic Magnetic Field}, \emph{Astrophys. J.} {\bf 761} (2012) L11. [arXiv:1210.7820]

\bibitem{Sun8}
X.~H. Sun, W. Reich, W. Waelkens and T.~A. En{\ss}lin, \emph{Radio observational constraints on Galactic 3D-emission models}, \emph{Astron. Astrophys.} {\bf 477} (2008) 573. [arXiv:0711.1572]

\bibitem{Sun10}
X.~H. Sun and W. Reich, \emph{The Galactic halo magnetic field revisited}, \emph{Res. Astron. Astrophys.} {\bf 10} (2010) 1287. [arXiv:1010.4394]

\bibitem{Gaia}
Gaia Collaboration, \emph{The Gaia mission}, \emph{Astron. Astrophys.} {\bf 595} (2016) A1. [arXiv:1609.04153]

\bibitem{Dust}
B-G Andersson, A. Lazarian and J. E. Vaillancourt, \emph{Interstellar Dust Grain Alignment}, \emph{Annu. Rev. Astron. Astrophys.} {\bf 53} (2015) 501.

\bibitem{PASIPHAE}
PASIPHAE website: \url{http://pasiphae.science/}.

\bibitem{PASIPHAE_paper}
K. Tassis et al., \emph{PASIPHAE: a high-galactic-latitude, high-accuracy optopolarimetric survey}, (2018) [arXiv:1810.05652]

\bibitem{Davis}
L. Davis, \emph{The Strength of Interstellar Magnetic Fields}, \emph{Phys. Rev.} {\bf 81} (1951) 890.

\bibitem{ChF}
S. Chandrasekhar and E. Fermi, \emph{Magnetic Fields in Spiral Arms}, \emph{Astrophys. J.} {\bf 118} (1953) 113. 

\bibitem{Hildebrand}
R.~H. Hildebrand \textit{et al.}, \emph{ Dispersion of Magnetic Fields in Molecular Clouds. I}, \emph{Astrophys. J.} {\bf 696} (2009) 567. [arXiv:0811.0813]



\bibitem{Panopoulou}
G.~V. Panopoulou, I. Psaradaki and K. Tassis, \emph{The magnetic field and dust filaments in the Polaris Flare}, \emph{Mon. Notices Royal Astron. Soc.} {\bf 462} (2016) 1517.

\bibitem{Soam_Pattle}
A. Soam, K. Pattle, D. Ward-Thompson et al., \emph{Magnetic Fields toward Ophiuchus-B Derived from SCUBA-2 Polarization Measurements}, \emph{Astrophys. J.} {\bf 861} (2018) 65.

\bibitem{Liu}
T. Liu, P.~S. Li, M. Juvela et al., \emph{A Holistic Perspective on the Dynamics of G035.39-00.33: The Interplay between Gas and Magnetic Fields}, \emph{Astrophys. J.} {\bf 859} (2018) 151.

\bibitem{Beuther}
H. Beuther, J.~D. Soler, W. Vlemmings et al., \emph{Magnetic fields at the onset of high-mass star formation}, \emph{Astron. Astrophys.} {\bf 614} (2018) A64.

\bibitem{Kwon}
J. Kwon, Y. Doi, M. Tamura et al., \emph{A First Look at BISTRO Observations of the $\rho$ Oph-A core}, \emph{Astrophys. J.} {\bf 859} (2018) 4.

\bibitem{Mao}
S.~A. Mao, B.~M. Gaensler, S. Stanimirovi{\'c} et al., \emph{A Radio and Optical Polarization Study of the Magnetic Field in the Small Magellanic Cloud}, \emph{Astrophys. J.} {\bf 688} (2008) 1029.
 
\bibitem{Soam_Lee}
A. Soam, C.~W. Lee, G. Maheswar et al., \emph{Probing the magnetic fields in L1415 and L1389}, \emph{Mon. Notices Royal Astron. Soc.} {\bf 464} (2017) 2403.

\bibitem{Clemens}
D.~P. Clemens, K. Tassis and P.~F. Goldsmith, \emph{The Magnetic Field of L1544. I. Near-infrared Polarimetry and the Non-uniform Envelope}, \emph{Astrophys. J.} {\bf 833} (2016) 176.

\bibitem{Golup}
G. Golup, D. Harrari, S. Mollerach and E. Roulet, \emph{Source position reconstruction and constraints on the galactic magnetic field from ultra-high energy cosmic rays}, \emph{Astropart. Phys.} {\bf 32} (2009) 269. 
[arXiv:0902.1742]

\bibitem{Keivani_diss}
A. Keivani, \emph{Magnetic deflections of ultra-high energy cosmic rays from Centaurus A}, \emph{LSU Doctoral Dissertations} (2013) 2426.
\url{ https://digitalcommons.lsu.edu/gradschool_dissertations/2426} 

\bibitem{CRT}
M. S. Sutherland, B. M. Baughman and J. J. Beatty, \emph{CRT: A numerical tool for propagating ultra-high energy cosmic rays through Galactic magnetic field models}, \emph{Astropart. Phys.} {\bf 34} (2010) 198. [arXiv:1010.3172]

\bibitem{uncertainty}
M. Unger and G. R. Farrar, \emph{Uncertainties in the Magnetic Field of the Milky Way} (2017). [arXiv:1707.02339]

\bibitem{Deflections}
G. R. Farrar and M. S. Sutherland, \emph{Deflections of UHECRs in the Galactic magnetic field} (2017) [arXiv:1711.02730]

\bibitem{Deflections_Centaurus}
A. Keivani, G. R. Farrar and M. Sutherland, \emph{Magnetic deflections of ultra-high energy cosmic rays from Centaurus A}, \emph{Astropart. Phys.} {\bf 61} (2015) 47.[arXiv:1406.5249]

\bibitem{Dir_stat}
K. V. Mardia and P. E. Jupp, \emph{Directional Statistics}, John Wiley \& Sons (2009), pg. 36.

\bibitem{von_Mises}
D. J. Best and N.I. Fisher, \emph{Efficient Simulation of the von Mises Distribution}, \emph{Appl. Statist.} {\bf 28} (1979) 152.


\end{thebibliography}
\end{document}